\documentclass[aps,prb,superscriptaddress,reprint,floatfix,showpacs]{revtex4-1}
\usepackage{dcolumn}
\usepackage{bm}
\usepackage{natbib}
\usepackage{longtable}
\usepackage{amsmath}
\usepackage{url}
\usepackage{multirow}
\usepackage[dvipdfmx]{graphicx}
\begin{document}

\title{Thickness-dependent electronic and magnetic properties of
$\bf{\gamma}'$-Fe$_{4}$N atomic layers on Cu(001)}

\author{Y. Takahashi}
\affiliation{Institute for Solid State Physics, The University of Tokyo,
Kashiwa, Chiba 277-8581, Japan}

\author{T. Miyamachi}
\email{toshio.miyamachi@issp.u-tokyo.ac.jp}
\affiliation{Institute for Solid State Physics, The University of Tokyo,
Kashiwa, Chiba 277-8581, Japan}

\author{S. Nakashima}
\affiliation{Institute for Solid State Physics, The University of Tokyo,
Kashiwa, Chiba 277-8581, Japan}

\author{N. Kawamura}
\affiliation{Institute for Solid State Physics, The University of Tokyo,
Kashiwa, Chiba 277-8581, Japan}
\affiliation{Science \& Technology Research Laboratories, NHK, Setagaya,
Tokyo 157-8510, Japan}

\author{Y. Takagi}
\affiliation{Department of Materials Molecular Science, Institute for Molecular Science, Myodaiji-cho, Okazaki 444-8585, Japan}
\affiliation{Department of Structural Molecular Science, The Graduate University for Advanced Studies (SOKENDAI), Myodaiji-cho, Okazaki 444-8585, Japan}

\author{M. Uozumi}
\affiliation{Department of Materials Molecular Science, Institute for Molecular Science, Myodaiji-cho, Okazaki 444-8585, Japan}
\affiliation{Department of Structural Molecular Science, The Graduate
University for Advanced Studies (SOKENDAI), Myodaiji-cho, Okazaki
444-8585, Japan}

\author{V. Antonov}
\affiliation{Max-Planck-Institut f\"ur Mikrostrukturphysik, Weinberg 2, 06120 Halle, Germany}
\affiliation{Institute for Metal Physics, 36 Vernadsky Street, 03142 Kiev, Ukraine}

\author{T. Yokoyama}
\affiliation{Department of Materials Molecular Science, Institute for Molecular Science, Myodaiji-cho, Okazaki 444-8585, Japan}
\affiliation{Department of Structural Molecular Science, The Graduate
University for Advanced Studies (SOKENDAI), Myodaiji-cho, Okazaki 444-8585, Japan}

\author{A. Ernst}
\affiliation{Max-Planck-Institut f\"ur Mikrostrukturphysik, Weinberg 2,
06120 Halle, Germany}

\author{F. Komori}
\email{komori@issp.u-tokyo.ac.jp}
\affiliation{Institute for Solid State Physics, The University of Tokyo, Kashiwa, Chiba 277-8581, Japan}

 \begin{abstract}
  Growth, electronic and magnetic properties of
  $\bf{\gamma}'$-Fe$_{4}$N atomic layers on Cu(001) are studied by scanning tunneling microscopy/spectroscopy and x-ray
  absorption spectroscopy/magnetic circular dichroism. A continuous film
  of ordered trilayer $\bf{\gamma}'$-Fe$_{4}$N is obtained by Fe deposition under
  N$_{2}$ atmosphere onto monolayer Fe$_{2}$N/Cu(001), while the repetition of
  a bombardment with 0.5 keV N$^{+}$ ions during growth cycles results in imperfect bilayer
  $\bf{\gamma}'$-Fe$_{4}$N. The increase in the sample thickness causes the
  change of the surface electronic
  structure, as well as the enhancement in the spin magnetic moment of Fe
  atoms reaching $\sim 1.4\ \mu_{\rm B}$/atom in the trilayer
  sample. The observed thickness-dependent properties of the system are well interpreted
  by layer-resolved density of states calculated using first principles,
  which demonstrates the strongly layer-dependent electronic states within
  each surface, subsurface, and interfacial plane of the $\bf{\gamma}'$-Fe$_{4}$N atomic
  layers on Cu(001).
 \end{abstract}


\maketitle

\section{Introduction}
\indent Iron nitrides, especially in iron-rich phases, have been under intense
research due to the strong ferromagnetism and interest in its physical origin
\cite{Coey1999Magnetic-nitrid,Frazer1958Magnetic-Struct}. The
difficulty in obtaining a single phase has been a long-standing problem
for ferromagnetic iron nitrides, to hinder fundamental understanding of
intrinsic physical properties \cite{Coey1994The-magnetizati,Komuro1990Epitaxial-growt,Ortiz1994Epitaxial-Fe16N}. Recently,
the successful epitaxial growth of single-phase ferromagnetic
$\bf{\gamma}'$-Fe$_{4}$N has been reported on various substrates, which
helps to comprehend a crucial role for the hybridization between Fe and
N states in the ferromagnetism of $\bf{\gamma}'$-Fe$_{4}$N
\cite{Atiq2008Effect-of-epita,Borsa2001High-quality-ep,Gallego2004Mechanisms-of-e,Ito2011Spin-and-orbita,Nikolaev2003Structural-and-,Kokado2006Theoretical-ana,Ito2015Local-electroni}. The
robust Fe-N bonding also renders an Fe$_{2}$N layer strongly
two-dimensional \cite{Fang2014Predicted-stabi}, which possibly
facilitates a layer-by-layer stacking of $\bf{\gamma}'$-Fe$_{4}$N on metals. This contrasts
with the case of elemental 3$d$ transition metals (TMs) deposited on 3$d$ TM substrates, in which inevitable atom
intermixing and exchange of constituents prevent the formation of ordered
overlayers \cite{Kim1997Subsurface-grow,Nouvertne1999Atomic-exchange,Torelli2003Surface-alloyin}. Therefore,
the investigation into the electronic and magnetic states of
$\bf{\gamma}'$-Fe$_{4}$N atomic layers can not only elucidate the
layer-/site-selective electronic and magnetic states of
$\bf{\gamma}'$-Fe$_{4}$N, but unravel the origin of the strongly thickness-dependent physical properties in a thin-film
limit of 3$d$ TM ferromagnets \cite{Srivastava1997Modifications-o,Farle1997Anomalous-reori,Farle1997Higher-order-ma,Schulz1994Crossover-from-,Li1994Magnetic-phases,Straub1996Surface-Magneti,Weber1996Structural-rela,Meyerheim2009New-Model-for-M}.\\
\indent Here, we report two growth modes of
$\bf{\gamma}'$-Fe$_{4}$N/Cu(001) depending on preparation methods. The scanning tunneling
microscopy/spectroscopy (STM/STS) observations indicated a successful
growth of ordered trilayer $\bf{\gamma}'$-Fe$_{4}$N, without extra
nitrogen bombardment onto the existing structures. X-ray absorption
spectroscopy/magnetic circular dichroism (XAS/XMCD) measurements
revealed the thickness dependence of the magnetic moments of Fe atoms,
the origin of which was well explained by the
first-principles calculations. Based on an atomically-resolved
structural characterization of the system, the layer-by-layer electronic and
magnetic states of the $\bf{\gamma}'$-Fe$_{4}$N atomic layers have been
understood from both experimental and theoretical points of view.\\
\section{Methods}
\indent A clean Cu(001) surface was prepared by repetition of sputtering
with Ar$^{+}$ ions and subsequent annealing at 820 K. Iron was deposited at room temperature (RT) in a preparation chamber under an ultrahigh vacuum (UHV) condition
($<1.0\times10^{-10}$ Torr), using an electron-bombardment-type
evaporator (EFM, FOCUS) from a high-purity Fe rod
(99.998 \%). The STM measurements were
performed at 77 K in UHV ($<3.0\times10^{-11}$ Torr) using
electrochemically etched W tips. The differential conductance d$I$/d$V$ was recorded for STS using a
lock-in technique with a bias-voltage modulation of 20 mV and 719
Hz. The XAS and XMCD measurements were
performed at BL 4B of UVSOR-I\hspace{-.1em}I\hspace{-.1em}I \cite{Gejo2003Angle-resolved-,Nakagawa2008Enhancements-of} in a total
electron yield (TEY) mode. The degree of circular
polarization was $\sim 65\ \%$, and the x-ray propagation vector lay within the (1\=10) plane of a Cu(001) substrate. All the
XAS/XMCD spectra were recorded at
$\sim8\ {\rm K}$, with external magnetic field $B$ up to $\pm5$ T
applied parallel to the incident
x-ray. The symmetry and quality of the surface were also checked by low energy electron
diffraction (LEED) in each preparation chamber. First-principles calculations were performed within the density
functional theory in the local density approximation
\cite{Perdew1992Accurate-and-si}, using a self-consistent full-potential Green function method specially
designed for surfaces and
interfaces \cite{Luders2001Ab-initio-angle,Geilhufe2015Numerical-solut}.\\
\section{Results and Discussion}
\subsection{\label{secbilayer}Monolayer and bilayer-dot $\bf{\gamma}'$-F\lowercase{e$_{4}$}N}
\begin{figure}
 \includegraphics[width=86mm]{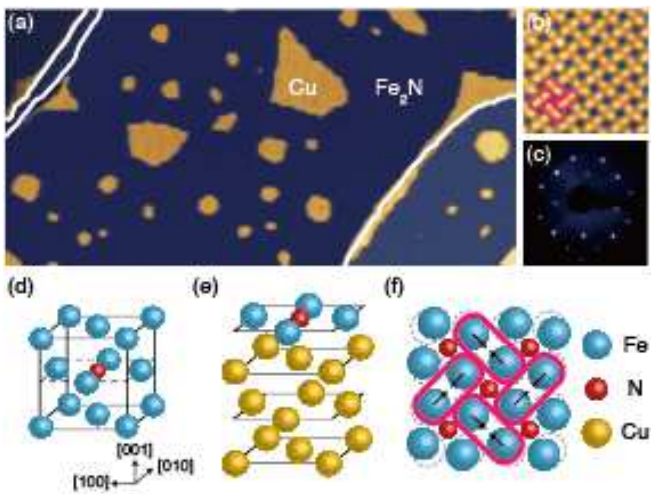}
 \caption{\label{fig1} (Color online) Topography and atomic structure of the monolayer
 $\bf{\gamma}'$-Fe$_{4}$N on Cu(001). (a) Topographic image (100$\times$50 nm$^2$,
 sample bias $V_{\rm s}=+1.0\ {\rm V}$, tunneling current $I=0.1\ {\rm nA}$)
 of the monolayer $\bf{\gamma}'$-Fe$_{4}$N on Cu(001). White lines represent
 step edges of the Cu(001)
 terraces. Color contrast is enhanced within each terrace. (b) Close view (2.5$\times$2.5 nm$^2$, $V_{\rm s}=0.25\ {\rm V}$, $I=45\ {\rm
 nA}$) of the surface Fe$_{2}$N layer. The dimerization of Fe atoms is indicated by encirclement. (c) LEED pattern obtained with an incident electron energy of 100
 eV. (d) Bulk crystal structure of $\bf{\gamma}'$-Fe$_{4}$N. A dotted parallelogram represents an Fe$_{2}$N plane. (e) Atomic structure of the
 monolayer $\bf{\gamma}'$-Fe$_{4}$N on Cu(001). (f) Schema illustrating $p4g(2\times2)$ reconstruction in the surface
 Fe$_{2}$N layer of $\bf{\gamma}'$-Fe$_{4}$N. Arrows indicate the shift of the Fe atoms from an
 unreconstructed $c(2\times2)$ coordination (dotted circles). For (d) to (f), large blue
 (yellow) and small red spheres represent Fe (Cu) and N
 atoms, respectively.}
\end{figure}
\indent Monolayer Fe$_{2}$N on Cu(001) was prepared prior to any growth
of multilayer $\bf{\gamma}'$-Fe$_{4}$N by the following cycle: N$^{+}$ ion
bombardment with an energy of 0.5 keV to a clean Cu(001) surface,
subsequent Fe deposition at RT, and annealing at 600 K. Note that the
monolayer Fe$_{2}$N is identical to Fe$_{4}$N on Cu(001) in a monolayer
limit, and thus referred to as also ''monolayer
$\bf{\gamma}'$-Fe$_{4}$N'' hereafter. A topographic image of the sample after one growth cycle is shown in
Fig. \ref{fig1}(a). The monolayer $\bf{\gamma}'$-Fe$_{4}$N is formed on the
Cu terraces at $\sim$ 0.85 ML coverage. An atomically-resolved
image of that surface displayed in
Fig. \ref{fig1}(b) reveals a clear
dimerization of the Fe atoms, typical of ordered
$\bf{\gamma}'$-Fe$_{4}$N on Cu(001)
\cite{Gallego20051D-Lattice-Dist,Takahashi2016Orbital-Selecti}. A LEED
pattern of the surface is shown in
Fig. \ref{fig1}(c), which exhibits sharp spots with the corresponding
$p4g(2\times2)$ symmetry. It is known that
\cite{Gallego20051D-Lattice-Dist,Gallego2004Self-assembled-,Navio2007Electronic-stru,Takahashi2016Orbital-Selecti} the topmost layer of the $\bf{\gamma}'$-Fe$_{4}$N on
Cu(001) always consists of the Fe$_{2}$N plane in a bulk Fe$_{4}$N crystal
shown in Fig. \ref{fig1}(d). A schematic model of the monolayer
$\bf{\gamma}'$-Fe$_{4}$N is given in Fig. \ref{fig1}(e), composed of
a single Fe$_{2}$N plane on Cu(001). Accordingly, the surface Fe$_{2}$N plane takes
reconstruction to the $p4g(2\times2)$ coordination \cite{Gallego20051D-Lattice-Dist}, in which the Fe
atoms dimerize in two perpendicular directions as illustrated in Fig. \ref{fig1}(f).\\
\begin{figure}
 \includegraphics[width=86mm]{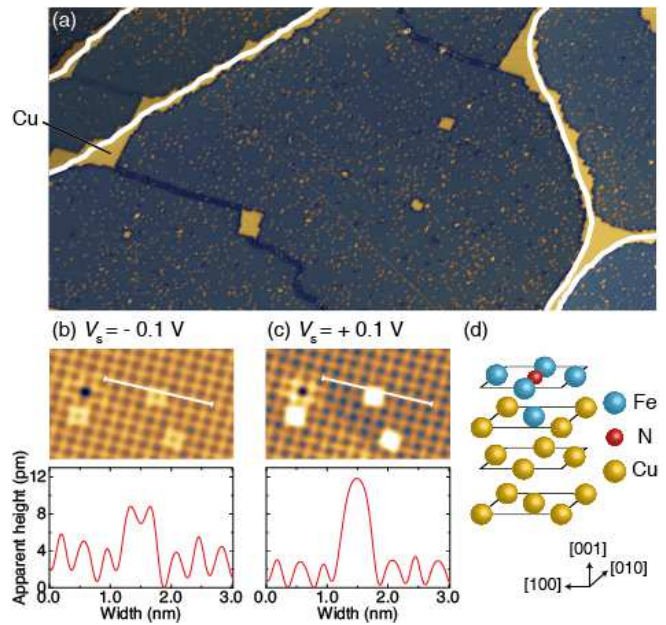}
 \caption{\label{fig2} (Color online) Topography of the bilayer
 $\bf{\gamma}'$-Fe$_{4}$N dot on Cu(001). (a) Topographic image (120$\times$60
 nm$^2$, $V_{\rm s}=-0.1\ {\rm V}$, $I=0.1\ {\rm nA}$)
 of the monolayer (darker area) and dot-like bilayer $\bf{\gamma}'$-Fe$_{4}$N
 on Cu(001). White lines represent step edges of the Cu(001)
 terraces. Color contrast is enhanced within each terrace. (b,c) Upper panels:
 Atomically-resolved topographic images (7$\times$3 nm$^2$, $I=2.0\ {\rm
 nA}$) taken at (b) $V_{\rm s}=-0.1\ {\rm V}$ and (c) $+0.1\ {\rm V}$. Lower panels:
 Height profiles measured along lines indicated in the upper panels. (d)
 Proposed atomic structure of the bilayer-dot $\bf{\gamma}'$-Fe$_{4}$N on
 Cu(001). Large blue
 (yellow) and small red spheres correspond to Fe (Cu) and N
 atoms, respectively.}
\end{figure}
\indent After repeating the growth cycles, we found a new structure different from the monolayer
$\bf{\gamma}'$-Fe$_{4}$N. Figure \ref{fig2}(a) displays the surface after two
growth cycles in total, namely, another cycle of the N$^{+}$ ion
bombardment, Fe deposition, and annealing onto the existing
monolayer $\bf{\gamma}'$-Fe$_{4}$N surface. Then,
the surface
becomes mostly covered with the monolayer $\bf{\gamma}'$-Fe$_{4}$N, which contains a small
number of bright dots. For a structural
identification of these dots, we measured atomically-resolved topographic
images and line profiles at different $V_{\rm
s}$ as shown in Fig. \ref{fig2}(b) and
\ref{fig2}(c). The dot structure imaged at $V_{\rm
s}=-0.1\ V$ reveals the dimerization of the Fe atoms as the monolayer
$\bf{\gamma}'$-Fe$_{4}$N surface. This indicates that the topmost part of the dot
consists of the reconstructed Fe$_{2}$N. At positive $V_{\rm s}$ of +0.1 V, in contrast, the dot is
recognized as a single protrusion both in the topographic image and line
profile, while the surrounding monolayer $\bf{\gamma}'$-Fe$_{4}$N still shows
the Fe dimerization. This implies the different electronic structure of the dot
compared to the monolayer $\bf{\gamma}'$-Fe$_{4}$N, which comes from
the difference in a subsurface atomic structure.\\
\indent The observed height difference between the dot and the monolayer
$\bf{\gamma}'$-Fe$_{4}$N ranges from 4 to 10 pm depending on $V_{\rm
s}$. These values are in the same order of a lattice
mismatch between the bulk crystals of the $\bf{\gamma}'$-Fe$_{4}$N/Cu(001)
(380 pm) and Cu(001) (362 pm) \cite{Gallego20051D-Lattice-Dist}, but an order of magnitude smaller than the lattice constant of
the $\bf{\gamma}'$-Fe$_{4}$N/Cu(001). This
suggests that the topmost layer of the dot is not located above the
monolayer $\bf{\gamma}'$-Fe$_{4}$N surface, but shares the Fe$_{2}$N
plane with. Furthermore, the bright
dot is composed of only four pairs of the Fe dimer as imaged in
Fig. \ref{fig2}(b), indicating that the difference in the atomic and/or
electronic structures is restricted within a small area. Considering the
above, it is most plausible that one Fe atom is
embedded just under the surface N atom at the dot center, and thus a bilayer
$\bf{\gamma}'$-Fe$_{4}$N dot is formed as
schematically shown in Fig. \ref{fig2}(d). This structure corresponds to a
minimum unit of the bilayer $\bf{\gamma}'$-Fe$_{4}$N on Cu(001).\\
\begin{figure}
 \includegraphics[width=86mm]{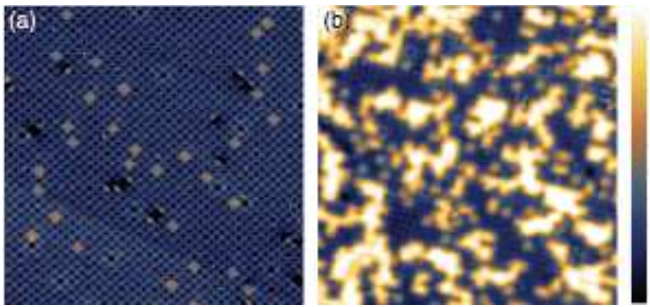}
 \caption{\label{fig3} (Color online) Topographic images (15$\times$15
 nm$^{2}$) of the surface
 after repetition of (a) two and (b) three growth cycles. The set point
 is ($V_{\rm s},\ I$) = (+0.25 V, 5.0 nA) for (a) and (+0.1 V, 3.0 nA) for (b).}
\end{figure}
\indent This bilayer dot formed clusters by a further
repetition of the growth cycles. Figure \ref{fig3}(a) shows an enlarged
view of the iron-nitride surface after two growth cycles. The
coverage of the dot is estimated to be $\sim$ 5 \% of
the entire surface. Another growth cycle onto this surface led to an
increase in a dot density up to $\sim$ 40 \%, as shown in
Fig. \ref{fig3}(b). However, further repetitions of the cycles resulted in neither a
considerable increase in the dot density nor the formation of a continuous bilayer film. This can be attributed to an inevitable sputtering
effect in every growth cycle: an additional N$^{+}$
ion bombardment to the existing surface not only implanted
N$^{+}$ ions but also sputtered the surface, which caused the loss
of the iron nitrides already formed at the surface, as well as the increase in the
surface roughness.\\
\indent To compensate this loss of surface Fe atoms by the sputtering
effect, we also tried to increase the amount of deposited Fe per
cycle. Nonetheless, the number of Fe atoms, which remained at the surface after annealing, did not
increase possibly because of the
thermal metastability of Fe/Cu systems
\cite{Detzel1994Substrate-diffu,Memmel1994Growth-structur,Shen1995Surface-alloyin,Bayreuther1993Proceedings-of-}. The
isolated Fe atoms without any bonding to N atoms were easily
diffused and embedded into the
Cu substrate during the annealing process. As a result, only the imperfect bilayer $\bf{\gamma}'$-Fe$_{4}$N was obtained
through this method.\\
\subsection{\label{sectrilayer}Trilayer $\bf{\gamma}'$-F\lowercase{e$_{4}$}N film}
\begin{figure}
 \includegraphics[width=86mm]{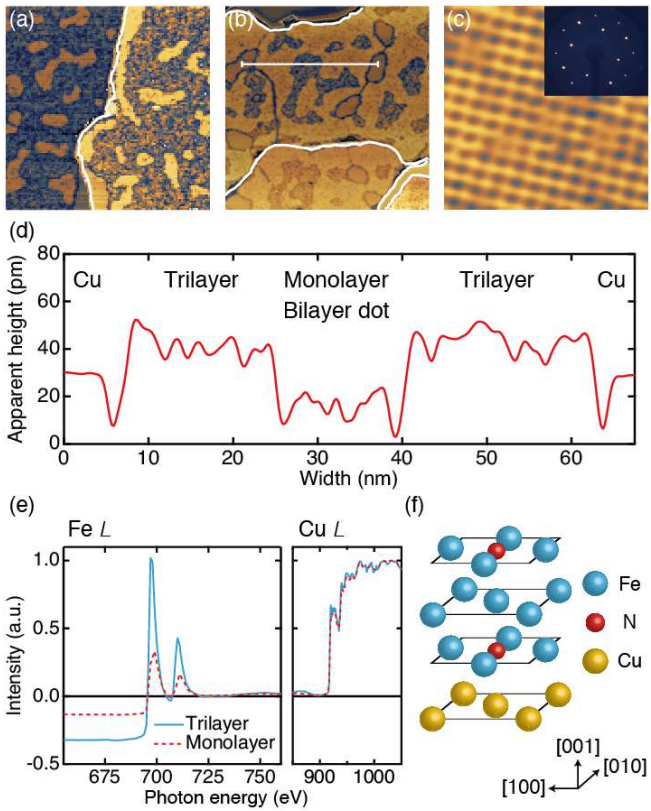}
 \caption{\label{fig4} (Color online) Topography of the trilayer $\bf{\gamma}'$-Fe$_{4}$N film on Cu(001). Topographic
 images (100$\times$100 nm$^2$) after (a) two and (b) three cycles of the Fe
 deposition under N$_{2}$ atmosphere and subsequent annealing onto the
 monolayer $\bf{\gamma}'$-Fe$_{4}$N on Cu(001). The setpoint is $I=0.1\
 {\rm nA}$, $V_{\rm s}=-0.1\ {\rm
 V}$ for (a) and -0.05 V for (b). White lines indicate step edges of the Cu terraces. Color contrast is enhanced within each terrace. (c) Atomically-resolved topographic image
 (4$\times$4 nm$^2$, $I=5.0\ {\rm nA},\ V_{\rm s}=-0.1\ {\rm V}$) of the
 trilayer $\bf{\gamma}'$-Fe$_{4}$N surface. An
 inset represents a LEED pattern of the sample shown in (b), obtained with
 an incident electron energy of 100 eV. (d) Height profile measured along
 the line indicated in (b). (e) XAS edge jump spectra of the trilayer
 (solid) and monolayer (dotted) samples at the Fe and Cu
 $L$ edges. The intensity is normalized to the Cu edge jump. (f) Atomic model expected for the trilayer $\bf{\gamma}'$-Fe$_{4}$N on Cu(001). Blue (yellow) large and red
 small spheres represent Fe (Cu) and N atoms, respectively.}
\end{figure}
\indent Multilayer $\bf{\gamma}'$-Fe$_{4}$N films were obtained by the following procedure. First, the monolayer
$\bf{\gamma}'$-Fe$_{4}$N was
prepared on Cu(001) as above. Then, 2 ML Fe was deposited under N$_{2}$ atmosphere (5.0$\times$10$^{-8}$ Torr) \footnote{We checked the ionization of nitrogen molecules/atoms without bombardment
using an ion gun. The ion flux monitored for the Fe evaporator increased in proportion to the rise in the N$_{2}$ pressure, far below
the parameters at which Fe started to be evaporated. This indicates the
ionization of the N$_{2}$ molecules and/or N atoms around the evaporator possibly by thermal electrons
created inside it. Then, the N$^{+}$ and N$_{2}^{+}$ ions could reach to the surface
together with the evaporated Fe atoms, or iron nitride was
already formed before landing.} at RT, and the sample was annealed at
600 K. Figures \ref{fig4}(a) and \ref{fig4}(b) show topographic images
after two and three above mentioned cycles, respectively. In the images, the coverage
of new bright area, different from the imperfect bilayer dot,
monotonously increases with repeating the cycles. A close view of that
new surface is displayed in Fig. \ref{fig4}(c), revealing the dimerized
(or even $c(2\times2)$-like dot) structures. Because a LEED pattern shown in the inset of Fig. \ref{fig4}(c)
exhibits the $p4g(2\times2)$
symmetry without extra spots, the topmost layer of this surface is
composed of the reconstructed Fe$_{2}$N plane
\cite{Takahashi2016Orbital-Selecti}. Therefore, these observations suggest that the
new area would consist of $\bf{\gamma}'$-Fe$_{4}$N other than both of the monolayer and bilayer dot.\\
\indent In order to determine the structure of this newly obtained
$\bf{\gamma}'$-Fe$_{4}$N, a typical height profile of the surface was recorded as shown in
Fig. \ref{fig4}(d). It is clear that the new structure is higher than both the
Cu surface and the surface including the monolayer/dot-like bilayer
$\bf{\gamma}'$-Fe$_{4}$N. This suggests that the new area is composed of
$\bf{\gamma}'$-Fe$_{4}$N thicker than bilayer. Quantitative information on the
thickness of the new structure could be obtained from Fe $L\
(2p\rightarrow3d)$ edge jump spectra shown in Fig. \ref{fig4}(e),
whose intensity is roughly proportional to the amount of
surface/subsurface Fe atoms. The sample
prepared in the same procedure as that shown in Fig. \ref{fig4}(b) reveals an edge jump value of 0.32, while the
monolayer $\bf{\gamma}'$-Fe$_{4}$N 0.12 \footnote{The amount of the Fe
atoms detected in the edge-jump spectra was smaller than that expected
from the initially deposited ones. This implies that a certain amount of Fe
atoms, not participating in forming any $\bf{\gamma}'$-Fe$_{4}$N structures,
was embedded into the Cu substrate during annealing, at least several nms
(probing depth in the TEY mode) below the surface.}. Considering that the new area
occupies $\sim$ 60 \% of the entire surface as deduced from
Fig. \ref{fig4}(b), the thickness of this $\bf{\gamma}'$-Fe$_{4}$N must be less than
quadlayer to meet the experimental edge jump value of 0.32 (See Appendix
\ref{jumptoML}). Hence, the newly
obtained structure is identified as a trilayer $\bf{\gamma}'$-Fe$_{4}$N
film. An atomic structure expected for
the trilayer $\bf{\gamma}'$-Fe$_{4}$N on Cu(001) is presented in Fig. \ref{fig4}(f). The
growth without any ion bombardment to the monolayer surface possibly
stabilizes the subsurface pure Fe layer, which
could promote the formation of the
trilayer $\bf{\gamma}'$-Fe$_{4}$N film in a large area.\\
\indent Finally, let us mention another growth method of the
$\bf{\gamma}'$-Fe$_{4}$N film. We previously report a possible layer-by-layer growth of the $\bf{\gamma}'$-Fe$_{4}$N atomic layers on
Cu(001), by the N$^{+}$ ion bombardment with a relatively low
energy of 0.15 kV \cite{Takagi2010Structure-and-m}. This soft
implantation of N$^{+}$ ions successfully
avoids extra damage to the existing $\bf{\gamma}'$-Fe$_{4}$N structures
during the repetition of the growth cycles. The reported different electronic/magnetic states could then originate from the difference in the fabrication processes. Another
finding is that, in the current study, only the monolayer and trilayer $\bf{\gamma}'$-Fe$_{4}$N
could be obtained in a continuous film form. This
implies that an Fe$_{2}$N-layer termination would be preferable through the present methods, possibly due to
the metastability of an interface between Cu and pure Fe layers \cite{Detzel1994Substrate-diffu,Memmel1994Growth-structur,Shen1995Surface-alloyin,Bayreuther1993Proceedings-of-}.\\
\subsection{\label{secelemag}Electronic and magnetic properties of $\bf{\gamma}'$-F\lowercase{e$_{4}$}N atomic layers}
\begin{figure}
\includegraphics[width=60mm]{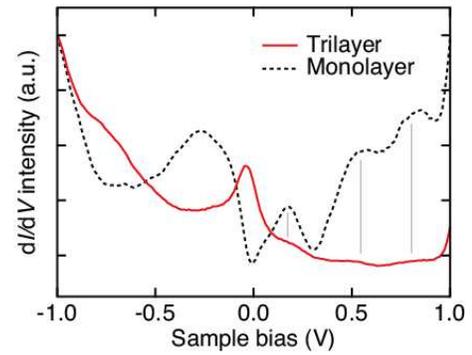}
 \caption{\label{fig5} (Color online) Surface electronic structures of
 the $\bf{\gamma}'$-Fe$_{4}$N on Cu(001). Experimental d$I$/d$V$
 spectra recorded above the trilayer (solid) and monolayer (dotted)
 $\bf{\gamma}'$-Fe$_{4}$N surfaces are presented. The d$I$/d$V$ intensity is arbitrary. A STM
 tip was stabilized at $V_{\rm s}=+1.0\ {\rm V}$, $I=3.0$ and 7.0 nA for
 the trilayer and monolayer surfaces, respectively. Gray lines are
 guide to the eye.}
\end{figure}
\indent The surface
electronic structures of $\bf{\gamma}'$-Fe$_{4}$N showed large dependence on
the sample thickness. Figure \ref{fig5} displays experimental d$I$/d$V$
spectra measured on the surfaces of the
trilayer and monolayer $\bf{\gamma}'$-Fe$_{4}$N. The
peaks located at $V_{\rm s}\sim$ +0.20, +0.55, and +0.80 V, mainly originating
from the unoccupied states in the down-spin band characteristic of
Fe local density of states (LDOS), are observed for both the trilayer and monolayer
surfaces. A significant
difference between the spectra is a dominant peak located around $V_{\rm
s}=-50\ {\rm mV}$ observed only for the trilayer surface. This peak possibly originates
from the LDOS peak located around $E-E_{\rm F}=-0.2\ {\rm eV}$, calculated
for the Fe atoms not bonded to N atoms in the subsurface Fe layer
[corresponding site of Fe4 shown in Fig. \ref{fig7}(b)]. Because of the $d_{\rm 3z^2-r^2}$
orbital character, this peak could be dominantly detected in the STS
spectrum for the trilayer surface. Thus, the appearance of this additional peak could support the different subsurface structure of the trilayer sample,
especially, the existence of the subsurface Fe layer proposed
above.\\
\begin{figure*}
\includegraphics[width=178mm]{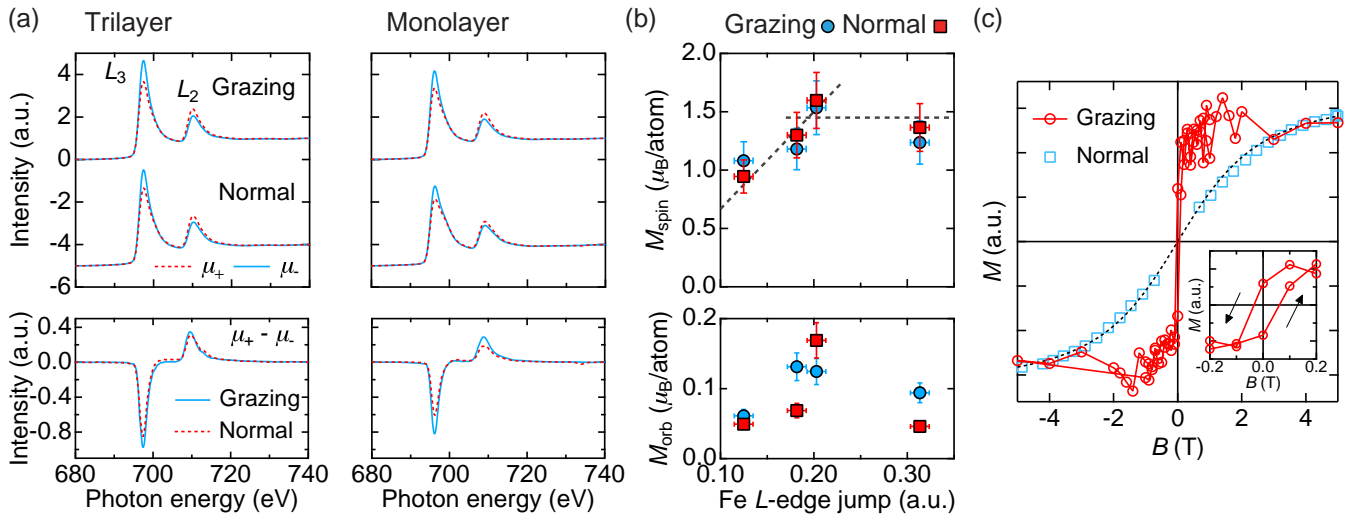}
 \caption{\label{fig6} (Color online) Thickness-dependent electronic
 and magnetic
 properties of the $\bf{\gamma}'$-Fe$_{4}$N atomic layers on Cu(001). (a) Upper panels: XAS spectra under $B=\pm5\ {\rm T}$ of the trilayer (left) and monolayer (right) samples in the grazing (top) and normal
 (bottom) incidence. Lower panels: Corresponding XMCD spectra in the grazing (solid) and normal
 (dotted) incidence. All the
 spectra are normalized to the Fe XAS $L$-edge
 jump. (b) Upper [lower] panel: Experimental spin [orbital] magnetic moment in the grazing (circle) and normal
 (square) incidence plotted with respect to the Fe $L$-edge jump values. The edge jump values of 0.12 and 0.32 correspond to
 those of the monolayer and trilayer samples, respectively. Dotted lines are
 guide to the eye. Error bars are indicated to all the data, and smaller
 than the marker size if not seen. (c) Magnetization of the monolayer sample recorded in the grazing (circle and line)
 and normal (square) incidence. A dotted line is the guide to
 the eye. An inset shows an enlarged view of the curve recorded in the
 grazing incidence.}
\end{figure*}
\indent The entire electronic and magnetic
properties of the sample, including both surface and subsurface information, were investigated by using XAS and XMCD techniques at the
Fe $L_{2,3}\ (2p_{1/2,3/2}\rightarrow 3d)$ absorption edges. Figure
\ref{fig6}(a) shows XAS ($\mu_{+},\ \mu_{-}$) and XMCD
($\mu_{+}-\mu_{-}$) spectra under $B=\pm5\
T$ of the trilayer and monolayer samples in the grazing
($\theta=55^{\circ}$) and normal incidence ($\theta=0^{\circ}$). Here, $\mu_{+}\ (\mu_{-})$
denotes a x-ray absorption spectrum with the photon helicity parallel (antiparallel) to the Fe 3$d$
majority spin, and an incident angle $\theta$ is
defined as that between the sample normal and incident x-ray. The
trilayer (monolayer) sample was prepared in the same procedure as that
shown in Fig. \ref{fig4}(b) [Fig. \ref{fig1}(a)]. It is clear that the XMCD intensity is larger in the trilayer
one, indicating an enhancement of magnetic moments with increasing
thickness.\\
\indent For a further quantitative analysis on the magnetic moments, we applied
XMCD sum rules \cite{Carra1993X-ray-circular-,Thole1992X-ray-circular-}
to the obtained spectra and estimated spin ($M_{\rm spin}$) and orbital
($M_{\rm orb}$) magnetic moments separately. Note that the average number of 3$d$
holes ($n_{\rm hole}$) of 3.2 was used in the sum-rule analysis, which
was estimated by comparing the area of the experimental XAS spectra with that of a reference spectrum of
bcc Fe/Cu(001) ($n_{\rm hole}=3.4$) \cite{Chen1995Experimental-Co}. The
thickness dependence of the $M_{\rm spin}$ and $M_{\rm orb}$ values is
summarized in Fig. \ref{fig6}(b). The value of $M_{\rm spin}$ increases
monotonously with increasing the Fe $L$-edge jump value, namely, an
average sample thickness, and finally saturates
at $\sim1.4\ \mu_{\rm B}$/atom in the trilayer sample (corresponding
edge jump value of 0.32). The change in $M_{\rm orb}$ is not so
systematic relative to $M_{\rm spin}$, however, the $M_{\rm orb}$ values
seem to be enhanced in the
grazing incidence. This implies
an in-plane easy magnetization of the $\bf{\gamma}'$-Fe$_{4}$N atomic layers on
Cu(001), also consistent with the previous reports on the
$\bf{\gamma}'$-Fe$_{4}$N thin films on Cu(001) \cite{Gallego2004Mechanisms-of-e,Takagi2010Structure-and-m}. Figure \ref{fig6}(c) shows magnetization curves of the
monolayer sample, whose intensity corresponds to the $L_{3}$-peak XAS intensity normalized to the $L_{2}$ one. The curve
recorded in the normal incidence shows negligible remanent
magnetization. On the other hand, that in the grazing one draws a
rectangular hysteresis loop, which confirms the in-plane easy magnetization. The coercivity
of the monolayer sample is estimated to be $\sim$ 0.05 T at 8.0 K, larger
 than $\sim$ 0.01 T for 5 ML Fe/Cu(001)
 \cite{Li1994Magnetic-phases}, $\sim$ 1 mT for 5 ML
 Fe/GaAs(100)-(4$\times$6) \cite{Xu1998Evolution-of-th} and the 30 nm
 thick $\bf{\gamma}'$-Fe$_{4}$N film \cite{Gallego2004Mechanisms-of-e} at RT.\\
\subsection{\label{sectheory}Theoretical analysis on the electronic and
  magnetic states of $\bf{\gamma}'$-F\lowercase{e$_{4}$}N atomic layers on Cu(001)}
\begin{figure}
\includegraphics[width=86mm]{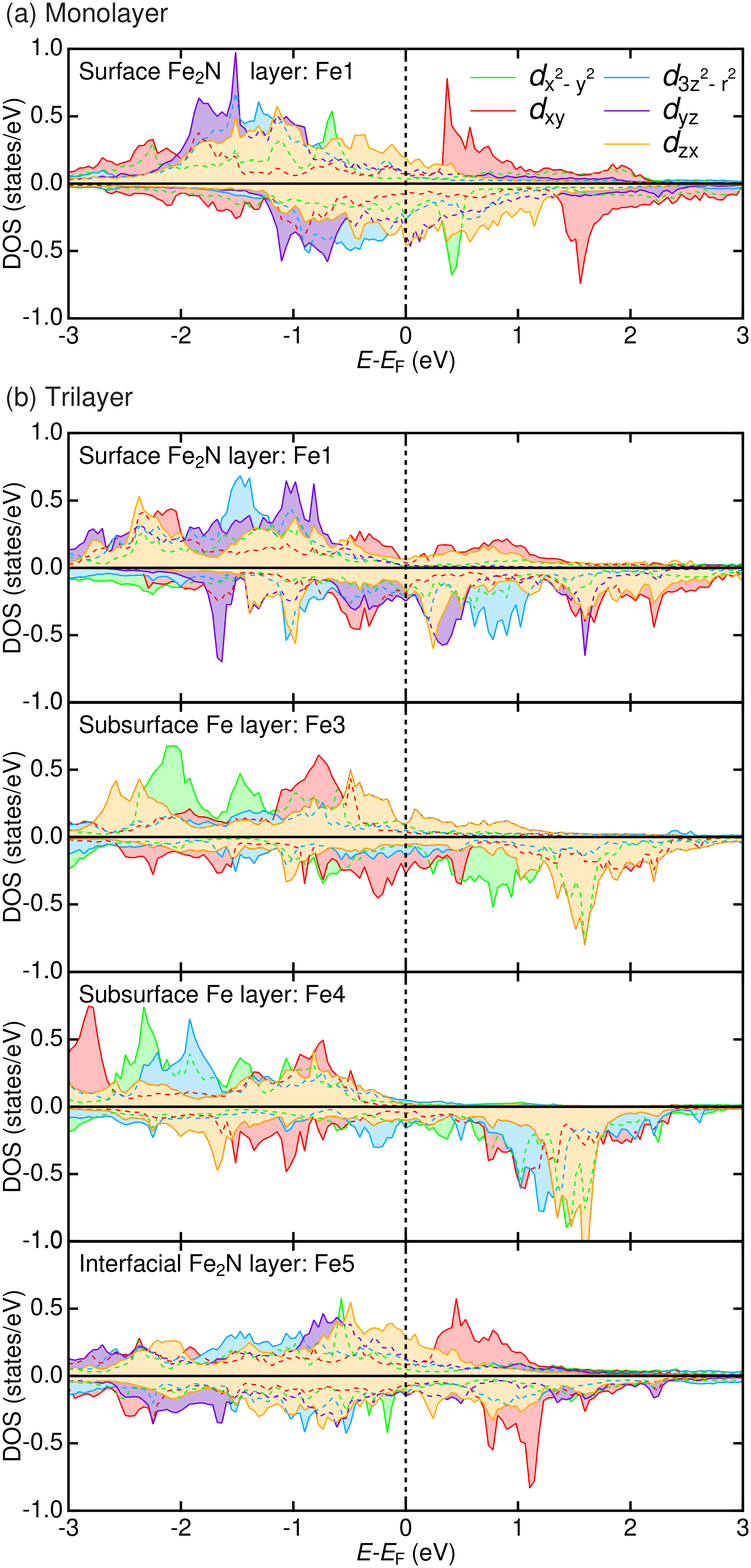}
 \caption{\label{fig7} (Color online) Layer-by-layer electronic states of the $\bf{\gamma}'$-Fe$_{4}$N atomic layers on Cu(001). Calculated layer-resolved DOS
 projected to each 3$d$ orbital of the (a) monolayer and (b) trilayer $\bf{\gamma}'$-Fe$_{4}$N on Cu(001). The DOS in the up-(down-)
 spin band is shown at upper (lower) panels. Note that the
 states with $d_{\rm yz}$ and $d_{\rm zx}$ orbitals are
 degenerated for the Fe3 and Fe4 sites in (b).}
\end{figure}
    \begin{table}
    \caption{\label{tab1} Calculated atomic magnetic
 moments of the Fe atoms at each site (in units of $\mu_{\rm
    B}$/atom). The site notation is the same as that used in Fig. \ref{fig7}.}
   \begin{tabular}{c|>{\centering\arraybackslash}p{3.0em}|>{\centering\arraybackslash}p{3.0em}|>{\centering\arraybackslash}p{3.25em}|>{\centering\arraybackslash}p{3.25em}|>{\centering\arraybackslash}p{3.5em}|>{\centering\arraybackslash}p{3.5em}}
    &\multicolumn{2}{>{\centering\arraybackslash}p{6.0em}|}{Surface Fe$_{2}$N}&\multicolumn{2}{>{\centering\arraybackslash}p{6.5em}|}{Subsurface Fe} &\multicolumn{2}{p{7.0em}}{Interfacial Fe$_{2}$N}\\
    & Fe1& Fe2 & Fe3 & Fe4 & Fe5 & Fe6\\ \hline
    Monolayer& 1.1 & 1.1 & -&-&-&-\\ \hline
    Trilayer& 1.8 & 1.8 & 2.0 & 3.0 & 0.62 &0.62
   \end{tabular}
   \end{table}
\indent The observed thickness dependence of the magnetic moments can be
well understood with a help of first-principles calculations. Figures \ref{fig7}(a)
and \ref{fig7}(b) show layer-resolved DOS of the monolayer and trilayer
$\bf{\gamma}'$-Fe$_{4}$N on Cu(001), respectively. Here, non-equivalent Fe sites in
 each layer are distinguished by different numbering. In particular, the Fe atoms at the Fe3 (Fe4) site in the trilayer
 $\bf{\gamma}'$-Fe$_{4}$N correspond to those with (without) a bond to
 N atoms \footnote{The difference of
DOS between (Fe1, Fe2) in the monolayer $\bf{\gamma}'$-Fe$_{4}$N, (Fe1,
Fe2) and (Fe5, Fe6) in the trilayer one is just a switch of the orbital
assignment between $d_{\rm yz}$ and $d_{\rm zx}$. Therefore, the DOS of
Fe2 in the monolayer $\bf{\gamma}'$-Fe$_{4}$N, Fe2 and Fe6 in the
trilayer one is not presented here.}. In Table \ref{tab1},
calculated values of an atomic magnetic moment $M_{\rm atom}$, corresponding to
$M_{\rm {spin}}$ + $M_{\rm {orb}}$ along the easy magnetization direction, are also listed. In the monolayer case, the
calculated $M_{\rm atom}$ is 1.1 $\mu_{\rm B}$/atom, which is in perfect
agreement with the experimental value. This supports an ideal
atomic structure of our monolayer sample.\\
\indent Interestingly, the value of $M_{\rm atom}$ for the Fe atoms in the
monolayer $\bf{\gamma}'$-Fe$_{4}$N is more than 1.5 times smaller than
that in the topmost layer of the
trilayer one (1.83 $\mu_{\rm B}$/atom). In comparison with the DOS shown at the top of
Fig. \ref{fig7}(b), the impact of the hybridization with the Cu states
on the Fe DOS can be seen in Fig. \ref{fig7}(a): First, the DOS in the
up-spin band, especially with $d_{\rm 3z^2-r^2}$ and $d_{\rm yz}$ orbitals, becomes
to have a tail toward a higher-energy side across the $E_{\rm F}$. This
change deviates the 3$d$ electrons in the up-spin band from a
fully-occupied nature. Moreover, the spin asymmetry of the occupied 3$d$
electrons, the difference between the electron occupation into each spin band
normalized by the sum of them, reduces especially for the DOS with
$d_{\rm xy}$, $d_{\rm 3z^2-r^2}$ and $d_{\rm yz}$ orbitals. These changes could decrease $M_{\rm
spin}$ of the Fe atoms. Note that the similar reduction in the magnetic
moments of 3$d$ TMs due to the hybridization with Cu states is
reported, for example, in
Ref. \onlinecite{Tersoff1982Magnetic-and-el,Hjortstam1996Calculated-spin}.\\
\indent Then, by comparing two different
Fe$_{2}$N interfaces with the Cu substrate, it turns out that $M_{\rm atom}$ of the
monolayer $\bf{\gamma}'$-Fe$_{4}$N (1.1 $\mu_{\rm B}$/atom) is almost twice compared to
that of the trilayer one (0.62 $\mu_{\rm B}$/atom). In the monolayer case, the Fe$_{2}$N layer faces to a vacuum and the
Fe atoms are under reduced atomic coordination. This results in the narrower
band width, and thus the DOS intensity increases in the
vicinity of $E_{\rm F}$. Accordingly, a larger exchange splitting can be
possible and the spin asymmetry of the occupied 3$d$ electrons increases as shown in Fig. \ref{fig7}(a), compared to the interfacial Fe$_{2}$N layer of the trilayer
$\bf{\gamma}'$-Fe$_{4}$N [bottom panel of Fig. \ref{fig7}(b)]. This leads to larger magnetic
moments at the surface. As a result, the competition between
the enhancement at the surface and the
decrease at the interface would make $M_{\rm atom}$ values quite layer-sensitive.\\
\indent In the subsurface Fe layer of
the trilayer $\bf{\gamma}'$-Fe$_{4}$N, the value of $M_{\rm atom}$ becomes
largest due to the bulk coordination of the Fe atoms. Especially the Fe atoms not
bonded to the N ones possess $M_{\rm atom}$ of 3.0 $\mu_{\rm B}$/atom,
which is comparable to the values of Fe atoms at the same site in the
bulk $\bf{\gamma}'$-Fe$_{4}$N \cite{Frazer1958Magnetic-Struct}. Consequently, by averaging the
layer-by-layer $M_{\rm atom}$ values of the trilayer
$\bf{\gamma}'$-Fe$_{4}$N, the total magnetic moment
detected in the XMCD measurement is expected to be 1.7 $\mu_{\rm
B}$/Fe, with the electron escape depth taken into account (See Appendix
\ref{jumptoML}). Considering the composition expected to the trilayer
sample, this value can well explain the experimental one of $\sim$ 1.5 $\mu_{\rm
B}$/Fe.\\
\indent The theory also demonstrates the direction of an easy magnetization
axis. The in-plane easy magnetization of
our $\bf{\gamma}'$-Fe$_{4}$N samples was confirmed by the magnetization
curves as well as the incidence
dependence of the $M_{\rm orb}$ value. In contrast, the pristine ultrathin Fe
films, which form either fct or fcc structures on Cu(001), show uncompensated
out-of-plane spins over a few surface layers
\cite{Pescia1987Magnetism-of-Ep,Meyerheim2009New-Model-for-M}. This shift
of magnetic anisotropy by nitridation can be understood from the orbital-resolved Fe DOS shown in Figs. \ref{fig7}(a) and \ref{fig7}(b). Unlike
the pure Fe/Cu(001) system \cite{Lorenz1996Magnetic-struct}, the occupation of 3$d$ electrons in states with out-of-plane-oriented orbitals ($d_{\rm yz},\
d_{\rm zx},\ d_{\rm 3z^2-r^2}$) is considerably larger than that with
in-plane-oriented ones ($d_{\rm
xy},\ d_{\rm x^2-y^2}$). This could make $M_{\rm orb}$ prefer to align
within a film plane, resulting in the in-plane magnetization of the system
\cite{Bruno1989Tight-binding-a}.\\
\section{Summary}
\indent In conclusion, we have conducted a detailed study on the growth,
electronic and magnetic properties of the $\bf{\gamma}'$-Fe$_{4}$N atomic layers on
Cu(001). The ordered trilayer film of $\bf{\gamma}'$-Fe$_{4}$N can be
prepared by the Fe deposition under N$_{2}$ atmosphere onto the existing monolayer surface. On the other hand, the repetition of
the growth cycles including the high-energy N$^{+}$ ion implantation
resulted in the imperfect bilayer $\bf{\gamma}'$-Fe$_{4}$N. The STM and STS observations revealed the change in the surface topography
and electronic structures with increasing the sample thickness. The XAS
and XMCD measurements also showed the thickness dependence of the spectra, and the corresponding
evolution of the $M_{\rm spin}$ values. All the thickness
dependence of the electronic and magnetic properties is well explained
by the layer-resolved DOS calculated using the first
principles. Structural perfection of the system makes it possible
to fully comprehend the layer-by-layer electronic/magnetic states of the $\bf{\gamma}'$-Fe$_{4}$N atomic layers.\\
\section{Acknowledgement}
\indent This work was partly supported by the JSPS Grant-in-Aid for Young Scientists (A), Grant No. 16H05963, for Scientific Research (B),
Grant No. 26287061, the Hoso Bunka Foundation, Shimadzu Science
Foundation, Iketani Science and Technology Foundation, and
Nanotechnology Platform Program (Molecule and Material Synthesis) of the
Ministry of Education, Culture, Sports, Science and Technology (MEXT),
Japan. Y. Takahashi was supported by
the Grant-in-Aid for JSPS Fellows and the Program for Leading Graduate
Schools (MERIT). A.E. acknowledges funding by the German Research
Foundation (DFG Grants No. ER 340/4-1).\\
\appendix
 \section{\label{jumptoML}Conversion of XAS edge jump values to the thickness of $\bf{\gamma}'$-Fe$_{4}$N}
 \indent The escape probability of electrons from inside a sample
 to a vacuum depends on the depth at which the electrons are
 excited. For a numerical interpretation of the XAS edge jump, the
 following factors
 should be mainly considered in principle: the penetration length of an
 incident x-ray ($\lambda_{x}$) and electron escape depth ($\lambda_{e}$), both
 energy-dependent. In the case of a few atomic layers of 3$d$
 transition metals, the attenuation of the incident
 x-ray intensity is almost negligible because $\lambda_{x}$ is orders of
 magnitude longer than the sample thickness
 \cite{Nakajima1999Electron-yield-}. Therefore, in the present case, only
 the electron escape probability at the depth $z$ from the
 surface, namely, a factor of $\exp{(-z/\lambda_{e})}$ is taken into
 account. As for the $\lambda_{e}$ value of Fe, 17 \AA\ was tentatively
 assumed in our analysis, which is experimentally determined for
 Fe thin films \cite{Nakajima1999Electron-yield-}. Then, based on the
 experimental Fe (N) edge
 jump values of 0.12 (0.015), those for the full-coverage dot-like
 bilayer, trilayer, and quadlayer $\bf{\gamma}'$-Fe$_{4}$N on Cu(001) are calculated as summarized in Table
 \ref{tablejump}.
  \begin{table}[h]
   \caption{\label{tablejump} Experimental and calculated Fe (N) edge
   jump values for the monolayer, dot-like bilayer, trilayer, and quadlayer $\bf{\gamma}'$-Fe$_{4}$N on Cu(001). In the
   calculation, each $\bf{\gamma}'$-Fe$_{4}$N is assumed to have the atomic
   structure presented in the text and fully cover the entire
   surface. For the quadlayer one, an
   Fe$_{2}$N/Fe$_{2}$/Fe$_{2}$N/Fe$_{2}$/Cu(001) structure is assumed.}
  \begin{tabular}{c|cc|cc}
   &\multicolumn{2}{c|}{Fe edge jump}&\multicolumn{2}{c}{N edge jump} \\
   &Experiment&Calculation&Experiment&Calculation \\ \hline
   Monolayer&\multicolumn{2}{c|}{0.12 (exp.)}&\multicolumn{2}{c}{0.015 (exp.)}\\
   Bilayer dot&\multicolumn{1}{c|}{-}&0.19&\multicolumn{1}{c|}{-}&0.015\\
   Trilayer&\multicolumn{1}{c|}{0.32}&0.40&\multicolumn{1}{c|}{0.032}&0.034\\
   Quadlayer&\multicolumn{1}{c|}{-}&0.57&\multicolumn{1}{c|}{-}&0.034
  \end{tabular}
  \end{table}
%
\end{document}